Fractional quantum Hall effect at Landau level filling $\nu=4/11$

W. Pan[1], K.W. Baldwin[2], K.W. West[2], L.N. Pfeiffer[2], and D.C. Tsui[2]

[1]Sandia National Labs, Albuquerque, New Mexico 87185, USA
[2]Princeton University, Princeton, New Jersey 08544, USA

We report low temperature electronic transport results on the fractional quantum Hall effect of composite fermions at Landau level filling $\nu = 4/11$ in a very high mobility and low density sample. Measurements were carried out at temperatures down to 15mK, where an activated magnetoresistance $R_{xx}$ and a quantized Hall resistance $R_{xy}$, within 1% of the expected value of $h/(4/11)e^2$, were observed. The temperature dependence of the $R_{xx}$ minimum at 4/11 yields an activation energy gap of ~ 7 mK. Developing Hall plateaus were also observed at the neighboring states at $\nu = 3/8$ and 5/13.

The search for novel fractional quantum Hall effect (FQHE) states continues to attract a great deal of interest since they were first discovered in 1982 [1]. Among many experimentally observed FQHE states, the even-denominator FQHE state in the second Landau level at Landau level filling $\nu=5/2$ [2,3] remains the most exotic one. There, conventional theories, such as the Laughlin wavefunction [4], hierarchical model [5,6], and weakly interacting composite fermion (CF) model [7,8,9], all fail to explain the origin of this FQHE state. Instead, a pairing mechanism of CFs [10] has to be evoked. Under this pairing picture, it is believed that the quasiparticles of this state obey the so-called non-Abelian statistics and, thus, may be useful in topological quantum computation [11].

To date, studies on the non-Abelian quantum Hall states have mostly been limited to the second Landau level. A little over ten years ago, the discovery of the FQHE at $\nu=4/11$ [12], however, generated new excitement on the existence of non-Abelian FQHE states in the lowest Landau level. This state has been viewed as an FQHE state of CFs. Yet, its origin remains elusive. Several proposals [13-18] have been made. Among them, the numerical simulations by Wójs, Yi, and Quinn (WYQ) [15] showed that a spin-polarized 4/11 FQHE state is an unconventional FQHE state of CFs and, possibly, a new non-Abelian state. Recently, using the CF diagonalization technique, Mukherjee et al [17] predicted a spin transition from a partially spin polarized state (which was believed to be the case for the experimentally observed 4/11 state [14]) to a fully spin-polarized state (or the WYQ state).

Despite a significant amount of theoretical work on this novel 4/11 FQHE state, experimentally, as correctly pointed out in Ref. [17], a definitive observation, in the form of an accurately quantized Hall plateau with activated longitudinal resistance, is still lacking. The difficulty is mainly due to the two conflicting requirements. Due to a limit in the highest field in a superconducting magnet, the electron density needs to be low and $\leq 1.6\times10^{11}$ cm$^{-2}$. On the other hand, the highest mobility is usually achieved at an electron density of $3\times10^{11}$ cm$^{-2}$, which would push the 4/11 state to a magnetic (B) field as high as 34T, much beyond the limit of a superconducting magnet. Recently, with new improvements in wafer growth, a high electron mobility of ~ $12\times10^6$ cm$^2$/vs has been achieved at an electron density of ~ $1.2\times10^{11}$ cm$^{-2}$. In this high-quality low-density sample, we observed at ν=4/11 activated magnetoresistance $R_{xx}$ and quantized Hall resistance $R_{xy}$, with the quantization better than 1%. Our results thus confirm that the 4/11 state is a true FQHE state. The temperature dependence of the $R_{xx}$ minimum at 4/11 yields an activation energy gap of ~7 mK.

The sample consists of a 50 nm wide modulation-doped GaAs/AlGaAs quantum well (QW) and has a size of about 5 mm × 5 mm. The QW is delta-doped with silicon from both sides at a distance of ~ 220 nm. Electrical contacts to the two-dimensional electron system (2DES) are accomplished by rapid thermal annealing of indium beats along the edge. The electron density of n = $1.17\times10^{11}$ cm$^{-2}$ and the mobility of μ = $11.6\times10^6$ cm$^2$/Vs were achieved after illumination of the sample at low temperatures by a red-light-emitting diode. A self-consistent calculation shows that at this density only one electric subband is occupied. All measurements were carried out in a dilution refrigerator with the lowest base temperature of ~ 15 mK. Low-frequency (~ 7Hz) lock-in amplifier techniques were utilized to measure $R_{xx}$ and $R_{xy}$. The excitation current was normally 2 or 5 nA.

Figure 1 shows the $R_{xx}$ and $R_{xy}$ traces in a large B magnetic field range from 5 to 14T. Well-developed FQHE are observed at ν = 2/3, 2/5, etc. Signatures of high-order FQHE states are observed up to 11/21 around 1/2, consistent with the ultra-high quality of this specimen.

In the field range between 1/3 and 2/5, similar to our previous work [12], developing FQHE states are observed at Landau level filling fractions ν = 4/11, 5/13, 3/8, and 6/17. Examining the Hall resistance, clearly, a plateau is formed at 4/11, with the quantization value of 2.725×h/e$^2$, within 1% of the expected value of 2.75×h/e$^2$ for the 4/11 state. Moreover, as shown in Fig. 3(a), the $R_{xx}$ at 4/11 displays an activated behavior. Its value increases with increasing temperatures. These two observations, quantized Hall resistance and activated magnetoresistance, confirm that indeed the 4/11 state is a true fractional quantum Hall effect. Developing Hall plateaus are also seen at filling factors 5/13 and 3/8. There is no visible feature in the Hall resistance around 6/17, consistent with a very weak minimum observed there. In fact, the 6/17 state is barely visible at the lowest temperature of 15 mK, as shown in Fig. 3(a). It becomes stronger as T increases. This observation, i.e., the appearance of a FQHE state at higher temperatures, has been observed before, and is believed to be due to the competition between a FQHE phase and a nearby insulating phase [19].

To further confirm the FQHE states at 4/11 and other filling factors, we compare in Fig. 2 $R_{xx}$ and $B \times dR_{xy}/dB$. Here $dR_{xy}/dB$ is the derivative of the $R_{xy}$ data with respect to B, obtained digitally from the $R_{xy}$ trace in Fig. 1. Overall, in the whole B field range, $R_{xx}$ and $B \times dR_{xy}/dB$ look very similar [20-22]. Strong minima are also seen in $B \times dR_{xy}/dB$ at $\nu=5/13$, 3/8, and 4/11. Moreover, the relative strength of the minima in $B \times dR_{xy}/dB$ also mimics that in $R_{xx}$. These observations from the comparison between $R_{xx}$ and $B \times dR_{xy}/dB$, again, are consistent with the FQHE states at $\nu=4/11$, 3/8, and 5/13.

Figure 3(a) shows $R_{xx}$ traces at three selected temperatures of 15, 22, and 30mK. It is clearly seen that the 4/11 state is activated. Its resistance increases with increasing temperatures. $R_{xx}$ at other filling factors $\nu=5/13$, 3/8, and 6/17, on the other hand, deceases with increasing temperatures, as seen in the past for fragile FQHE states (e.g., the 5/2 state [2]) when they were first observed. We believe that these states will eventually become activated with further improvement in sample quality. In Fig. 3(b), we show the $R_{xx}$ value at $\nu=4/11$ as a function of 1/T in a semi-log plot. From the linear fit to the data, though within a very limited range, an energy gap of ~ 7 mK is obtained.

We notice that the measured activation energy gap is much smaller than the numerical calculations, where the energy gap for a (partially) spin polarized 4/11 state has been estimated [14,17] to be (0.001) 0.002×$e^2/\varepsilon l_B$, or (0.18) 0.37 K. Here, e is the electron charge, ε the dielectric constant of GaAs, $l_B = (\hbar/eB)^{1/2}$ the magnetic length, ℏ the reduced Planck constant. This larger discrepancy is not unexpected and has been observed at many fragile FQHE states, for example at $\nu=5/2$ [3]. The exact origin of this large discrepancy is still under debate. Nevertheless, it is widely accepted that the sample disorder plays an important role. In order to estimate disorder broadening ($\Gamma$), we first use the so-called transport scattering time of ~ 440 ps, deduced from the zero-field mobility of $11.6\times10^6$ cm$^2$/Vs and effective mass of m* = 0.067m$_e$ (m$_e$ is the free electron mass). The so obtained $\Gamma$ is merely ~ 10 mK, much smaller than the theoretical calculated vales. On the other hand, if the quantum life time of ~ 8 ps, obtained from the on-set of Shubnikov-de Haas oscillations, is used, a disorder broadening $\Gamma$ ~ 0.5K is obtained, which is larger than the theoretically calculated ones. These two estimations show that the energy gap reduction at 4/11 is probably not related to either the transport scattering time or quantum life time of electrons. In view of this, we note that in a recent publication [23] the high-temperature resistance of the 5/2 state (at which the 5/2 state is supposed to be a Fermi sea state) was used as a criterion for judging the FQHE features in the second Landau level. Following this same line of thoughts, we calculate the disorder broadening using the CF transport scattering time, which was estimated to be ~ 50 ps. With this value, a disorder broadening of ~ 80 mK is obtained. This brings the theoretical values to ~ (0.1) 0.25K. The finite thickness of the 2DES in our sample will further reduce the energy gap to ~ (0.05) 0.12K. Further reduction of the theoretical gap due to Landau level mixing [24-33] is expected to bring the experimentally measured value closer to the numerical estimation.

Having established the 4/11 state a true FQHE state, in the following, we focus on the spin polarization of this exotic FQHE state. The spin polarization of the 4/11 state has been discussed in many numerical calculations [13-18]. In two recent publications [17,18], a spin transition from a two-component, partially spin polarized FQHE state (an Abelian state) to a single-component, spin-polarized FQHE state (or the WYQ state, possibly non-Abelian) is predicted to occur at $\kappa$ ~ 0.017-0.025, or in the B field range of ~ 9-19T. Here $\kappa = E_z/E_c$. $E_z$ ($\approx 0.3\times B$, in units of Kelvin) is the Zeeman energy and $E_c$ ($\approx 50.8\times B^{1/2}$, also in units of Kelvin) the Coulomb energy. Indication of such a spin transition was reported [34] by analyzing the resonant inelastic light scattering experiments in a thin quantum well sample of 33 nm with a low electron density of $5.5\times10^{10}$ cm$^{-2}$. There, a spin excitation mode below the Zeeman energy was observed at B ~ 7 T (or at the tilt angle of 30°). This mode disappeared when the sample was further tilted to 50° and the total B field ($B_{total}$) ~ 10 T at ν=4/11. This change in the character of the excitation was believed to be associated with a change in the spin polarization of the ground states in the Landau level filling range of 2/5 > ν > 1/3.

Back to our experiments, the indication of a spin-polarized 4/11 state [12] does not agree with the numerical simulation [14]. In fact, almost all the existing finite-size numerical calculations [13-18] seem to favor a partially spin polarized ground state for the experimentally observed 4/11 state. However, the numerical simulations were carried out in 2DES of zero thickness. On the other hand, it is known that finite thickness of 2DES reduces the critical Zeeman energy due to softening of the short-range interaction and the reduction of the interaction energy difference between different spin-polarization states [18]. Comparing to the sample used in the light scattering experiments, the samples in our electronic transport studies have a much wider quantum well, 50nm. Consequently, the critical Zeeman energy (or $\kappa_c$) at the spin transition is reduced, probably by half at $\lambda/l_B$ = 1.5 (the value estimated for our samples) [18,24]. Here λ is the effective width of the 2DES. This brings the transition point in the field range of ~5-10T. The 4/11 state in both samples we examined occurs at a B field higher than 10T. Consequently, it is expected to be spin-polarized. A spin polarized 4/11 state was further corroborated in a tilt magnetic field experiment. As reported in Ref. [12], the 4/11 state remains unchanged from zero tilt angle to ~ 40°, while the Zeeman energy increases by 1K. This increase in Zeeman energy is larger than the 4/11 energy gap and would have destroyed the 4/11 state if it is spin unpolarized or partially polarized.

To further support finite thickness of 2DES playing an important role in the spin polarization of the 4/11 state, we show in Fig. 4 the tilt magnetic field dependence of the 4/11 state in a thinner quantum well of 40 nm at T ~ 35 mK. The resistance minimum at 4/11 decreases with increasing tilt angle, different from that in the 50nm QW. Moreover, the relative $R_{xx}$ value of the 4/11 state, as defined in Fig. 4 and to take into account the background resistance due to tilting, also decreases. This is drastically different from the 50 nm sample, where the relative $R_{xx}$ value remains the same in the same field range from ~ 11.5 to 15T. It is possible that the 4/11 state in this 40 nm sample is very close to the spin transition point due to a lesser effect on the finite thickness correction. Increasing tilt angle helps push the 4/11 state deeper into the spin-polarized regime and thus strengthens this state.

Before we conclude this Rapid Communication, we note that the Landau level mixing effect can also affect the spin polarization of a FQHE ground state. The Landau level mixing effect is normally defined by $E_c/\hbar\omega_c$, where $\omega_c = eB/m^*$ is the cyclotron energy. It was observed recently [24] that, for the 3/5 state, $\kappa_c$ decreases linearly with increasing Landau level mixing effect and $\kappa_c$ is nearly halved when $E_c/\hbar\omega_c = 0.6$. In our sample, $E_c/\hbar\omega_c \sim 0.7$. Thus, the critical B field at the transition from the trivial two-component state to the WYQ state is expected to be further reduced to ~ 3-5T, again supporting a spin-polarized 4/11 state at B ~ 12T. However, extra scrutiny is needed here, since Landau level mixing itself affects significantly the ground state of a FQHE. It was reported in Ref. [24] that for the FQHE states around $\nu=3/2$ the Landau level mixing effect actually increases $\kappa_c$. This is unexpected and may have important implications on the 4/11 state. Under the CF picture, the 4/11 state is viewed as the 4/3 FQHE state of CFs in the second CF Landau level around the CF filling factor of 3/2. Following the above conclusion of the Landau level mixing effect on $\kappa_c$ for electrons, it is possible that a similar increase of $\kappa_c$ for the FQHE of CFs around the 3/2 filling may also occur. This can counter the finite thickness effect and move the critical magnetic field of the spin transition higher.

In summary, we studied the exotic FQHE state at Landau level filing factor $\nu=4/11$ in a ultra-high quality sample. We observed activated magnetoresistance $R_{xx}$ and quantized Hall resistance $R_{xy}$, within 1% of the expected value of $h/(4/11)e^2$. The temperature dependence of the $R_{xx}$ minimum at 4/11 yields an activation energy gap of ~ 7mK. Our results thus firmly establish that the 4/11 state is a true FQHE state. On the other hand, the nature of this FQHE state, in particular, the spin polarization of its ground state, remains enigmatic.

This work was supported by the U.S. Department of Energy, Office of Science, Basic Energy Sciences, Materials Sciences and Engineering Division. Sandia National Laboratories is a multi-program laboratory managed and operated by Sandia Corporation, a wholly owned subsidiary of Lockheed Martin Corporation, for the U.S. Department of Energy's National Nuclear Security Administration under Contract No. DE-AC04-94AL85000. A portion of this work was performed at the National High Magnetic Field Laboratory, which is supported by NSF Grant Nos. DMR-0084173 and *ECS-0348289*, the State of Florida, and DOE. Sample growth at Princeton was partially funded by the Gordon and Betty Moore Foundation as well as the National Science Foundation MRSEC Program through the Princeton Center for Complex Materials (DMR-0819860).

Note added in proof: After this paper was accepted for publication, we became aware of a paper by Liu *et al* [35] on possible FQHE of CFs at Landau level fillings 4/5 and 5/7 and work by Samkharadze *et al* [36] on the 4/11 and 5/13 states.

Figures and Figure Captions:

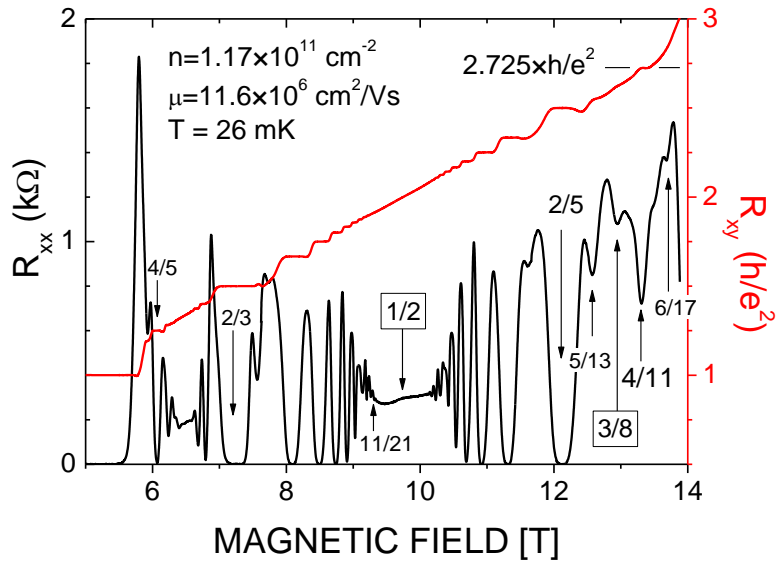

Figure 1 (color online) $R_{xx}$ and $R_{xy}$ in the magnetic field range of 5-14T at T = 26 mK. Representative fractions are marked by arrows.

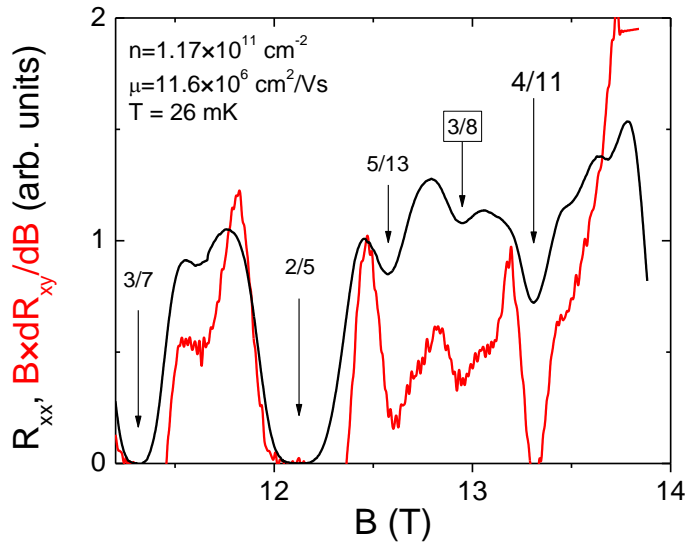

Figure 2 (color online) $R_{xx}$ and $B \times dR_{xy}/dB$ in the regime of $3/7 > \nu > 1/3$.

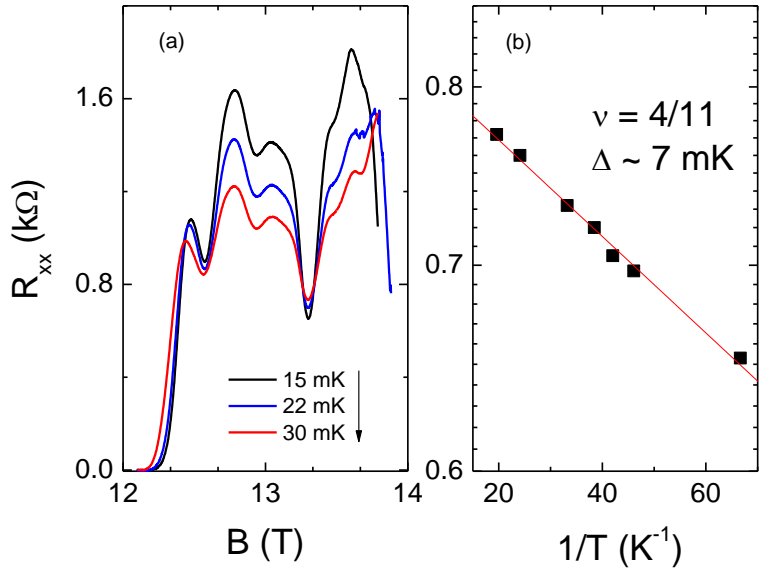

Figure 3 (color online) (a) T dependence of $R_{xx}$ between $2/5 > \nu > 1/3$. Three traces are shown at T = 15, 22, and 30 mK. (b) Arrhenius plot for the $R_{xx}$ minimum at $\nu=4/11$. The linear fit to the data points yields an energy gap of ~ 7 mK.

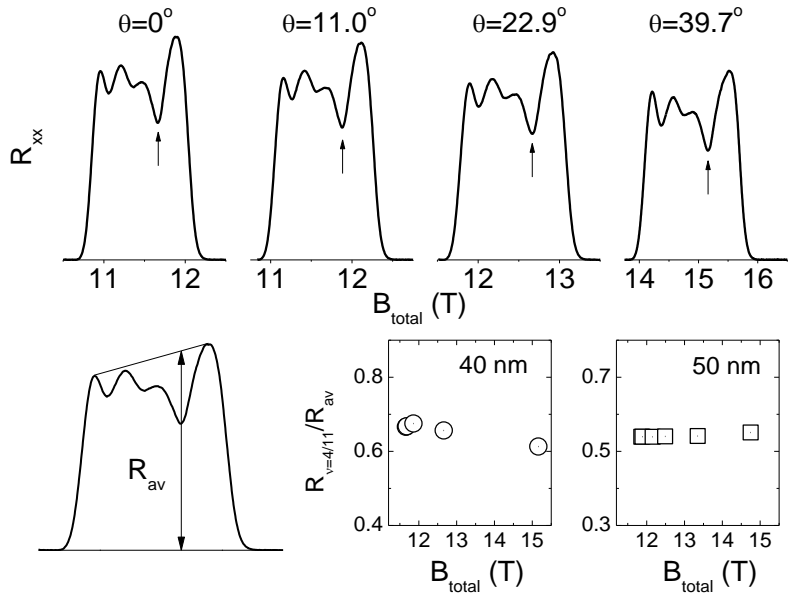

Figure 4 Top panel: $R_{xx}$ between $2/5 > \nu > 1/3$ at 4 selected tilt angles in a 40 nm width quantum well. The position of the 4/11 state is marked by arrow. Bottom left panel shows the definition of the background resistance $R_{av}$. The bottom right two panels show the relative $R_{xx}$ value of the 4/11 state $R_{\nu=4/11}/R_{av}$ *versus* the total B field ($B_{total}$) for the 40 and 50 nm QW samples, respectively.

11